# Electric and magnetic polarizabilities of hexagonal $Ln_2$CuTiO$_6$ ($Ln$=Y, Dy, Ho, Er and Yb)

Debraj Choudhury[1,2], Abhijit Hazarika[1], Adyam Venimadhav[3], Chandrasekhar Kakarla[3], Kris T. Delaney[4], P. Sujatha Devi[5], P. Mondal[5], R. Nirmala[1], J. Gopalakrishnan[1], Nicola A. Spaldin[4], Umesh V. Waghmare[6], D. D. Sarma[1,*]

[1]Solid State and Structural Chemistry Unit, Indian Institute of Science, Bangalore-560 012, India.
[2]Department of Physics, Indian Institute of Science, Bangalore-560012, India.
[3]Cryogenic Engineering Centre, Indian Institute of Technology, Kharagpur-721302, India.
[4]University of California, Santa Barbara, CA 93106-5050, U.S.A.
[5]Central Glass and Ceramic Research Institute, Kolkata-700032, India.
[6]Jawaharlal Nehru Centre for Advanced Scientific Research, Bangalore-560064, India.

We investigated the rare-earth transition metal oxide series, $Ln_2$CuTiO$_6$ ($Ln$=Y, Dy, Ho, Er and Yb), crystallizing in the hexagonal structure with non-centrosymmetric P6$_3$cm space group for possible occurrences of multiferroic properties. Our results show that while these compounds, except $Ln$=Y, exhibit a low temperature antiferromagnetic transition due to the ordering of the rare-earth moments, the expected ferroelectric transition is frustrated by the large size difference between Cu and Ti at the B-site. Interestingly, this leads these compounds to attain a rare and unique combination of desirable paraelectric properties with high dielectric constants, low losses and weak temperature and frequency dependencies. First-principles calculations establish these exceptional properties result from a combination of two effects. A significant difference in the $MO_5$ polyhedral sizes for $M$ = Cu and $M$ = Ti suppress the expected co-operative tilt pattern of these polyhedra, required for the ferroelectric transition, leading to relatively large values of the dielectric constant for every compound investigated in this series. Additionally, it is shown that the majority contribution to the dielectric constant arises from *intermediate-frequency* polar vibrational modes, making it relatively stable against any temperature variation. Changes in the temperature stability of the dielectric constant amongst different members of this series are shown to arise from changes in relative contributions from soft polar modes.

# 1. INTRODUCTION

Multiferroicity, defined as the simultaneous presence of two ferroic order parameters (usually, spontaneous presence of magnetization and electric polarization), has attracted huge attention in recent times.[1,2] Relatively small number of known cases of multiferroic materials has been explained[3,4] in terms of contradictory requirements for magnetism and ferroelectricity. While magnetism requires formation of moments, typically provided by finite number of $d$ electrons at a transition metal site, proper ferroelectricity is associated with a polar off-centering of a transition metal ion favored by its $d^0$ state. In order to have a finite macroscopic electric polarization, the crystal structure has to necessarily be non-centrosymmetric. Thus, one obvious way to search for a multiferroic material is to explore non-centrosymmetric crystal structures with presence of two types of transition metal ions, one with a finite $d$-occupancy and the other with a $d^0$ configuration. In such a situation, magnetism and ferroelectricity have the possibility of being supported in the two different sub-lattices.[5,6] In the ideal situation, these two order parameters may also influence each other due to the inevitable presence of electronic coupling between the two sub-lattices via finite hopping interactions.[6] With this background in mind, we identified $Ln_2CuTiO_6$ ($Ln$=rare earth) crystallizing in the non-centrosymmetric P6$_3$cm structure and containing $Cu^{2+}$ ($3d^9$) and $Ti^{4+}$ ($3d^0$) ions as a likely candidate to exhibit multiferroic properties. Our experiments, however, establish that neither $Cu^{2+}$ ions nor $Ti^{4+}$ ions order magnetically or ferroelectrically. While $Ln_2CuTiO_6$ shows a low temperature (~4 K) antiferromagnetic transition, this arises from a magnetic ordering of the rare-earth, $Ln^{3+}$, sub-lattice; this is experimentally established by the observation that the Y-analogue, $Y_2CuTiO_6$, in absence of any $4f$-electrons of rare-earth ions, does not exhibit the magnetic transition. We ascribe the absence of any long-range ferroelectric order to the existence of extensive size mismatch between the different trigonal bipyramids surrounding the B-site cations in these compounds. Detailed first-principles calculations suggest that the expected ferroelectric ordering is indeed frustrated by such size disorders. Interestingly, this suppressed ferroelectricity in the structure gives rise to several desirable dielectric properties in this series of compounds, as reported[7] very recently by us for a single compound in this series, namely for $Ho_2CuTiO_6$. The present study, while generalising the earlier observation of unusual dielectric properties of a single compound to the entire family of compounds, probes the microscopic origin of these results by comparative theoretical and experimental investigations and also elucidates magnetic properties of these compounds.

## 2. METHODS

$Ln_2CuTiO_6$ compounds ($Ln$ = Y, Dy, Ho, Er, Yb) were prepared following the solid state synthesis procedure[8]. Stoichiometric amounts of $Ln_2O_3$, CuO and $TiO_2$ powders were thoroughly mixed and sintered at 900°C for 24 hours in air, followed by sintering at 1050°C for 150 hours in air in pellet forms with three intermittent grindings. Dielectric measurements were performed using Instek LCR-821 High Precision LCR Meter and Solartron Frequency Response Analyzer (FRA 1260) and the *P-E* loops were recorded using Radiant RT6000 Test System. Dielectric constant values, obtained separately with sputtered gold or silver paste as electrodes, were found to be same, thereby ruling out any electrode polarization contribution. Magnetizations of the samples were recorded as a function of temperature from 2 K to 300 K and magnetic fields up to 5 Tesla, using a SQUID magnetometer. First-principles calculations were carried out within density-functional theory with a generalized-gradient approximation (GGA)[9] using PAW potentials as implemented in VASP.[10] A Hubbard U-correction (U=5 eV), following Dudarev implementation[11], was included for the d-states of Cu and Ti to correct for GGA deficiencies for the localized *d* states. This correction also ensures that the material is insulating with a non-vanishing band gap. Periodic boundary conditions were employed with three formula units per unit cell (a 30-atom periodic unit). The use of periodic boundary conditions prevents simulation of a completely disordered alloy, but the dielectric properties, which are dominated by long wavelength phonons, are expected to vary weakly with such details. For calculations, the transition-metal cations were ordered with a proportion of 2:1 within a given c-plane (and 1:2 in alternate c-planes) such that the total composition remained $Cu_{0.5}Ti_{0.5}$. Vibrational frequencies and normal modes were determined using the frozen phonon method (with ionic displacements of ±0.04 Å) and were used along with nominal ionic charges to calculate both the IR active modes characterized by non-zero mode effective charges (dipole) and their contribution to the static dielectric response.

## 3. RESULTS AND DISCUSSIONS

$Ln_2CuTiO_6$ compounds ($Ln$ = Y, Dy, Ho, Er, Yb) were found to crystallize in the hexagonal space group $P6_3cm$. The structure consists of layers of corner shared ($Cu^{2+}$/$Ti^{4+}$)$O_5$ trigonal bipyramids alternating with hexagonal layers of sevenfold coordinated $Ln^{3+}$ along the $c$ axis (see Figure 1). It has been reported that $a$ (=$b$) and $c$ lattice parameters range between 6.095 Å- 6.214 Å and 11.482 Å – 11.515 Å, respectively.[8]

We show the magnetic susceptibilities [$\chi$] of two compounds, one with $Ln$ = Ho and the other with Y, as a function of temperature [$T$] in Figure 2. Both compounds show a Curie-Weiss behavior with $Ho_2CuTiO_6$ exhibiting an antiferromagnetic transition at $T_N$ = 4 K. The absence of this transition in $Y_2CuTiO_6$ suggests that this low temperature transition is associated with an ordering of the magnetic moments on the rare-earth site, while the $Cu^{2+}$ moments do not order.

The negative intercept (~ -161 K) of $\chi^{-1}$ vs. T plot for $Y_2CuTiO_6$ suggests overall antiferromagnetic interactions between $Cu^{2+}$ ions and the slope of the Curie-Weiss fit gives the effective magnetic moment of 2.1 Bohr magnetons ($\mu_B$) per $Cu^{2+}$ ion. Similar high value for negative temperature intercept (~ -330 K) from Curie –Weiss fit has also been observed for Mn ions in isostructural $YMnO_3$.[12] For a free ion $Cu^{2+}$, assuming quenched orbital angular momentum,[13] the calculated magnetic moment is only 1.7 Bohr magnetons. The larger experimental value of 2.1 suggests a finite orbital angular momentum contribution to the total angular momentum, as it adds in the same sense as the spin angular momentum for more than half filled $d^9$ ($Cu^{2+}$) system. In this context, we note that sizable angular orbital momentum can be found[14] even for a $Mn^{2+}$ $d^5$ oxide system, where nominally one would expect a zero angular momentum also in the atomic limit.

Both Ho and Cu contribute to the magnetic signal in $Ho_2CuTiO_6$, however, it is expected to be dominated by the much larger magnetic moment (10.6 $\mu_B$) of $Ho^{3+}$ compared to that of $Cu^{2+}$. The overall interaction is antiferromagnetic, as is seen from the negative intercept (~ -21 K) of $\chi^{-1}$ vs. $T$ plot. The $\chi$ vs. $T$ data is analysed with the help of a model, which assumes a linear superposition of independent antiferromagnetic correlations between the Ho spins and the Cu spins. This is consistent with an earlier observation[15] that susceptibility of $HoMnO_3$ could be understood as independent contributions from the Ho sublattice and the Mn sublattice. In order to analyze the experimental data, the effective magnetic moment (2.1 $\mu_B$) and antiferromagnetic interaction strength (-161 K) between copper spins are taken to be the same for $Ho_2CuTiO_6$ as in $Y_2CuTiO_6$, under the assumption that the two magnetic sublattices (Ho and Cu) do not influence each other and the $Cu^{2+}$ in $Ho_2CuTiO_6$ is very similar to that in $Y_2CuTiO_6$. The fitting of the experimental susceptibility data with this model gives the effective magnetic moment to be 11.4 $\mu_B$ per $Ho^{3+}$ and (-) 22 K as the antiferromagnetic interaction strength between them. The obtained value of the antiferromagnetic interaction strength (-22 K) between just the Ho spins is indeed similar to the overall interaction strength (~ -21 K) obtained from the intercept of the $\chi^{-1}$ vs. $T$ plot for $Ho_2CuTiO_6$. The free ion $Ho^{3+}$, in the ground state, has $S=2$, $L=6$ and $J=8$, which gives the effective magnetic moment as 10.6 $\mu_B$/ $Ho^{3+}$, being very close to the experimentally determined number of 11.4 $\mu_B$. We understand the absence of any magnetic ordering in the Cu sub-lattice by noting that the B-sub lattice in these compounds contains a disordered arrangement of the B-site ($Cu^{2+}$ and $Ti^{4+}$) cations, which would naturally suppress any long range magnetic ordering of the $Cu^{2+}$ spins.

Figure 3 shows the dielectric constant of $Er_2CuTiO_6$ as a function of temperature for several frequencies. It is evident that there is very little dependence of the dielectric constant on the temperature over a wide range of frequencies (100 Hz $\leq f \leq$ 100 kHz); the total variation in $\varepsilon'_r$ over this entire $T$ range (15 K $\leq T \leq$ 430 K) for frequencies between 100 Hz and 100 kHz is only 7%. Dielectric constant values at 1 MHz, measured for some temperatures, were found to be very similar to those measured at 100 kHz. Specifically, the temperature coefficient of the dielectric constant, $TC_{\varepsilon'}$ ($TC_{\varepsilon'}= (1/ \varepsilon'_r)\times(\partial \varepsilon'_r /\partial T)$) at 300 K, in this case is (-) 253 ppm $K^{-1}$ and (-) 263 ppm $K^{-1}$ at 100 Hz and 100 kHz. This compares favourably with the best value of (-)64 ppm $K^{-1}$ at 100 kHz reported recently for $Ho_2CuTiO_6$.[7] We also note that $\varepsilon'_r$ is uniformly large compared to most high-k materials, being discussed in the recent literature.[7] In order to compare the temperature dependence of $\varepsilon'_r$ for all compounds, studied here, we have plotted the normalized dielectric constants, $\varepsilon'_N = \varepsilon'_r(T) / \varepsilon'_r(300\ K)$, for all compounds in Figure 4 on an expanded scale as a function of T measured with a constant frequency of 100 kHz; the behavior is quite similar at other frequencies too. This plot clearly shows that $Ho_2CuTiO_6$ has the smallest dependence and $Yb_2CuTiO_6$ has the largest dependence of $\varepsilon'_r$ on T. In order to focus on the frequency dependence, we have plotted $\varepsilon'_r$ at 300 K as a function of the frequency for all compounds in Figure 5. From this plot, we find that the frequency dependence of the $Er_2CuTiO_6$ compound is the least, while it is the highest for the $Ho_2CuTiO_6$ compound. The extreme stability of dielectric constant values with changes in frequency for various $Ln_2CuTiO_6$ compounds are quantified by their frequency coefficients, given by the percentage variation of dielectric constants from their respective values at 1 kHz, over the frequency range between 100 Hz to 100 kHz, at 300K and lies between ±0.5% for $Er_2CuTiO_6$ to ±3.1% for $Ho_2CuTiO_6$. Finally we show $\varepsilon'_r(\omega)$ for a specific compound, namely $Er_2CuTiO_6$, at different temperatures in Figure 6. This plot suggests that the frequency dependence of $\varepsilon'_r$ is relatively temperature insensitive for these compounds. We have summarized the important dielectric properties of all these compounds in Table I. We note that all the $Ln_2CuTiO_6$ compounds possess impressively high dielectric constant values. Notably all of them are also characterized by extremely low loss values as seen in Table I, comparable to that (0.002) of $BaTi_4O_9$, a material known for its extremely low loss.[16] The small frequency coefficients for various $Ln_2CuTiO_6$ compounds are comparable to that of the gate dielectric material $HfO_2$, for which the frequency coefficient is ±1 %.[17]

It is interesting to note that the dielectric properties of all these compounds are similar without any obvious systematic trend reflecting the lanthanide contraction while the $Ln^{3+}$ ion is varied from Dy to Yb. This can be understood from the structural information.[8] The relevant structural parameters are listed in Table II. While the *a* and *b* cell parameters reflect the gradual lanthanide contraction, the *c* parameter is found to increase at the same time. The overall cell volume does reflect the lanthanide contraction. However, the most important structural parameters controlling dielectric properties are the transition metal (Cu/Ti) – oxygen bond distances. In this structure, there are two axial *M*-O distances and one equatorial one. These are listed in Table II; it is evident that these change very little from one compound to other and also do not exhibit any definite trend.

Uniformly large values of $\varepsilon'_r$ and a low loss (*D*), as shown in Table I, make these samples a new class of high-K materials with technological possibilities. These samples remain paraelectrics down to the lowest temperature measured (~10 K), as seen from the linear dependencies of their polarizations (P) as a function of the applied electric fields (E) without the observation of any P-E loop. Also, our zero temperature theoretical calculations show these samples to be paraelectric. Furthermore, we probed the possibility of any coupling between magnetic and dielectric properties in these compounds by carrying out magnetocapacitance measurements in the presence of a magnetic field of 5 Tesla and temperatures down to 10 K. Any possible effect of the antiferromagnetic transition, present around 4 K, on the dielectric properties, is expected to have an imprint even at ~ 10 K, this being just 6 K above the magnetic $T_N$. We find complete absence of any magnetocapacitative coupling in the entire temperature range as shown in Figure 7 for $Ho_2CuTiO_6$ ($T_N$ = 4 K), thus confirming the absence of any coupling between the magnetic and dielectric degrees of freedom in these compounds. It is, however, intriguing to find that in spite of having a large $\varepsilon'_r$ and possessing a non-centrosymmetric crystal structure, these samples remain paraelectrics. We have investigated this puzzling phenomenon with the help of first principles calculations.

We first note that the structure of $Ln_2CuTiO_6$ compounds is isomorphous with the prototypical hexagonal manganite, $YMnO_3$. There is an important difference, however: The B site in $YMnO_3$ is uniformly occupied by $Mn^{3+}$ ions, whereas the B sites in the $Ln_2CuTiO_6$ compounds are occupied by two different transition metal ions --$Ti^{4+}$ and $Cu^{2+}$ -- with vastly different oxidation states and ionic sizes. $YMnO_3$ has been extensively studied experimentally and theoretically,[18,19] and is a uniaxial ferroelectric with a polarization of ~6 $\mu C/cm^2$ along the hexagonal c axis. The ferroelectricity is unusual in that the primary order parameter for the structural transition from the paraelectric to ferroelectric phase is the long-range condensation of the K3 non-polar rotational mode of the $MnO_5$ trigonal bipyramids; the polar relative displacements of the Y and O ions, which cause the ferroelectric polarization, then couple linearly to this mode.[18,19] Similar physics for the origin of ferroelectricity has also been observed in isostructural $InMnO_3$.[20]

Since such rotational modes involve long-range correlations of the polyhedra -- if one polyhedron rotates to the left its neighbor must rotate by an equivalent amount to the right[18] -- they are easily frustrated by disorder. This has been demonstrated in the case of perovskite structure oxides where it has been shown to affect the competition between rotational and polar modes.[21]

To test whether disorder is also responsible for frustration of the rotational modes in these hexagonal systems, we calculated the stability of the K3 mode, and corresponding possible ferroelectricity, for $Ho_2CuTiO_6$ and $Y_2CuTiO_6$. Our calculations around the calculated ground states of the $Ln_2CuTiO_6$ compounds, using 30 atom super cells, which is compatible with the K3 mode, however, show that the K3 phonon mode is not unstable in these structures.[22] We attribute the stability to the significant difference of sizes of the trigonal bi-pyramids of $Cu^{2+}$ and $Ti^{4+}$. For example, in case of $Y_2CuTiO_6$, we find the volumes of $Ti^{4+}$ and $Cu^{2+}$ trigonal bi-pyramids to be 6.2 Å$^3$ and 7.3 Å$^3$ respectively. Similarly in case of $Ho_2CuTiO_6$, the $Ti^{4+}$ and $Cu^{2+}$ trigonal bi-pyramids are found to be 6.4 Å$^3$ and 7.6 Å$^3$ respectively. Since the K3 instability does not occur, the trigonal bipyramids remain completely aligned along the c-axis without any tilting, and the structures remain non-ferroelectric. This shows that ferroelectricity in these structures, though expected, can be very easily frustrated by such B-site size-mismatch that stabilizes the K3 mode. This effect is further enhanced by the presence of Cu-Ti anti-site disorder.

Besides large dielectric constants and low losses, this series of compounds is characterized by small values of $TC_{\varepsilon'}$ (see Table I), illustrating a robustness of the dielectric constant against changes in the temperature (Figure 4), with $Ho_2CuTiO_6$ and $Y_2CuTiO_6$ having the lowest and highest values of $TC_{\varepsilon'}$ within this series. At high frequencies, explored in our measurements, the dielectric response of a material has contributions from both electrons and phonons. The electronic contribution, which is the square of the refractive index, usually exhibits a weak dependence on temperature. Therefore, any strong temperature dependence of the dielectric constant typically arises from the phonon contribution, particularly when a material is close to a structural phase transition of polar or ferroelectric character. Such transitions are characterized by Γ-point vibrational modes which are both *soft* (below ~100 cm$^{-1}$) and polar. We have earlier shown[7] that the ionic dielectric response of Ho$_2$CuTiO$_6$ is dominated by intermediate frequency polar phonon modes, rather than the soft modes; since anharmonicities result in a strong temperature dependence of the softest modes, this was identified with the origin of the unusually low $TC_{\varepsilon'}$ in this series of compounds. In order to critically test this suggestion, we have now computed the frequency spectrum of the Γ-point phonon density of states for Y$_2$CuTiO$_6$, which has the highest $TC_{\varepsilon'}$ within this series of compounds in order to carry out a comparative study between Y$_2$CuTiO$_6$ and Ho$_2$CuTiO$_6$ with the lowest $TC_{\varepsilon'}$ in the group. The total phonon DOS at the Γ point along with those arising only from the polar modes are shown for Y$_2$CuTiO$_6$ in Figure 8a. A comparison with the corresponding quantities for Ho$_2$CuTiO$_6$ from ref 7 and shown as an inset to Figure 8a clearly demonstrate that Y$_2$CuTiO$_6$ has preponderance of softer polar phonon modes compared to those in Ho$_2$CuTiO$_6$. Contributions to dielectric constants from various polar modes are plotted as a function of the corresponding polar frequencies for Y$_2$CuTiO$_6$ and Ho$_2$CuTiO$_6$ in Figure 8b. These plots clearly show that there is a relatively larger contribution of soft phonon modes to the dielectric constant of Y$_2$CuTiO$_6$, consistent with its larger $TC_{\varepsilon'}$ value compared to those of Ho$_2$CuTiO$_6$.

In order to understand the chemical origin of such unusual dielectric responses, we further examine the eigenvectors of the softest IR active modes. We find that in Y$_2$CuTiO$_6$, as well as in Ho$_2$CuTiO$_6$, the softest modes involve small polar displacements of Cu. These being non-$d^0$ ions, contribute weakly to the dielectric response, and also exhibit a weak $T$-dependence in contrast to the soft modes in ferroelectrics where the $d^0$-ionic displacements in the soft modes cause ferroelectric transition and a strongly $T$-dependent dielectric response. We further find that the intermediate frequency IR active modes in Y$_2$CuTiO$_6$ involving Ti $d^0$ ionic displacements contribute strongly to its dielectric response.

## 4. SUMMARY

We illustrate that the $Ln_2$CuTiO$_6$ family of compounds ($Ln$ = Y, Dy, Ho, Er, Yb) order antiferromagnetically at low temperatures, except Y$_2$CuTiO$_6$, related to the ordering of the rare earth moments and they exhibit unusual combinations of large and extremely stable dielectric responses. Using first-principles calculations, we understand that size-disorder in the B-site arrangement of ions, present in all these compounds, is crucial for the origin of such unusual dielectric properties. We further find that the dielectric response in these compounds are dominated by intermediate frequency vibration modes, responsible for its extreme temperature stabilities. Our results show that use of B-site size-disorders in alloys of hexagonal transition metal oxides will generally lead to similar robust properties and may open up a new direction for designing technologically important dielectric materials.

## ACKNOWLEDGEMENTS

DDS, UW and DC acknowledge financial supports from DST and CSIR, Government of India. J.G. acknowledges financial support from the Indian National Science Academy. KTD and NAS acknowledge support from NSF under award DMR-0940420. Calculations were performed at the UCSB California Nanosystems Institute (CNSI) with facilities provided by NSF Award No. CHE-0321368 and Hewlett-Packard, at the San Diego Supercomputer Center, and at the National Center for Supercomputer Applications.

incommensurate cells adopt alternative (higher energy) cooperative tilt patterns that are still strongly polar. Note also that the K-point is mapped to Γ in our supercell. This means that if the rotational mode were unstable, it would be directly visible in the Γ – point phonon spectrum.

| Compound | $\varepsilon'_r$ | D | $TC_{\varepsilon'}$ (ppmK$^{-1}$) | Frequency Coefficient |
|---|---|---|---|---|
| Y$_2$CuTiO$_6$ | 40.3 | 0.003 | (-)713 | ±0.9% |
| Dy$_2$CuTiO$_6$ | 40.8 | 0.002 | (-)443 | ±0.6% |
| Ho$_2$CuTiO$_6$ | 53 | 0.005 | (-)64 | ±3.1% |
| Er$_2$CuTiO$_6$ | 40.1 | 0.003 | (-)263 | ±0.5% |
| Yb$_2$CuTiO$_6$ | 43.6 | 0.0017 | (-)564 | ±0.9% |

Table I: A comparison of the values of dielectric constants, losses (*D*), temperature coefficients (TC$_{\varepsilon'}$) of dielectric constants for various *Ln*$_2$CuTiO$_6$ compounds at 300 K and at frequency of 100 kHz. The values of frequency coefficients are given for a temperature of 300 K.

| Compound | a=b (Å) | c (Å) | Volume (Å$^3$) | d_axial | d_eq |
|---|---|---|---|---|---|
| Dy$_2$CuTiO$_6$ | 6.214 | 11.483 | 384.0 | 1.84, 1.72 | 2.07 |
| Ho$_2$CuTiO$_6$ | 6.180 | 11.499 | 380.3 | 1.84, 1.72 | 2.06 |
| Y$_2$CuTiO$_6$ | 6.172 | 11.482 | 378.8 | 1.84, 1.72 | 2.06 |
| Er$_2$CuTiO$_6$ | 6.144 | 11.506 | 376.1 | 1.84, 1.73 | 2.05 |
| Yb$_2$CuTiO$_6$ | 6.095 | 11.515 | 370.5 | 1.84, 1.73 | 2.03 |

Table II: A comparison of the cell parameters (*a*, *b* and *c*), cell volumes, two unequal axial and three equal equatorial transition metal - oxygen bond distances (given by d_axial and d_eq respectively) in *M*O$_5$ (*M* =Cu/Ti) trigonal bipyramids for various *Ln*$_2$CuTiO$_6$ compounds.

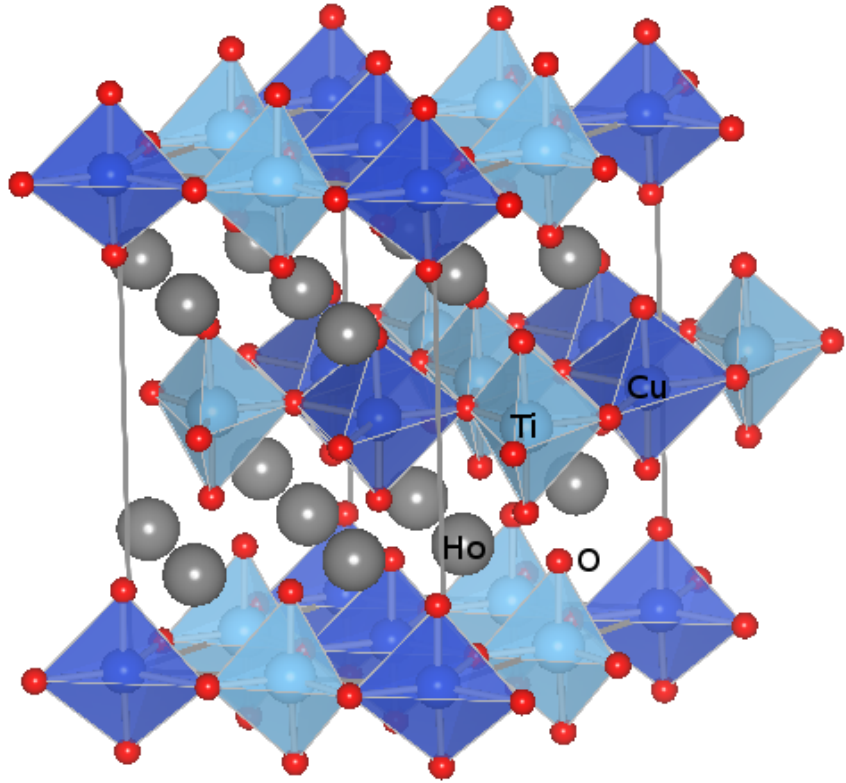

Figure 1: (Color Online) The hexagonal unit cell of $Ln_2CuTiO_6$ ($Ln$ = Y, Dy, Ho, Er, Yb) (structure shown for a representative member, $Ho_2CuTiO_6$). While $Ln_2CuTiO_6$ has disorder between the Cu and Ti sites, the crystal structure shown above consists of the transition metal ions ordered with a proportion of 1:2 and 2:1 in alternate $c$- planes respectively, with respect to which the calculations were performed.

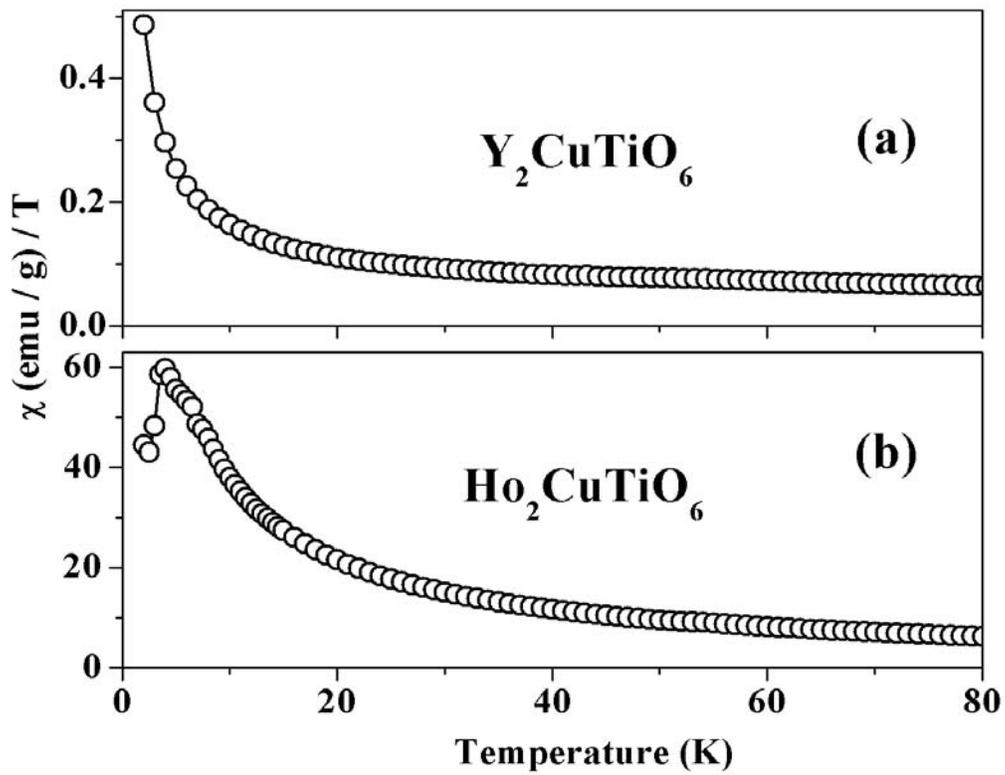

Figure 2: Temperature dependence of the susceptibility of (a) $Y_2CuTiO_6$ and (b) $Ho_2CuTiO_6$ as a function of temperature.

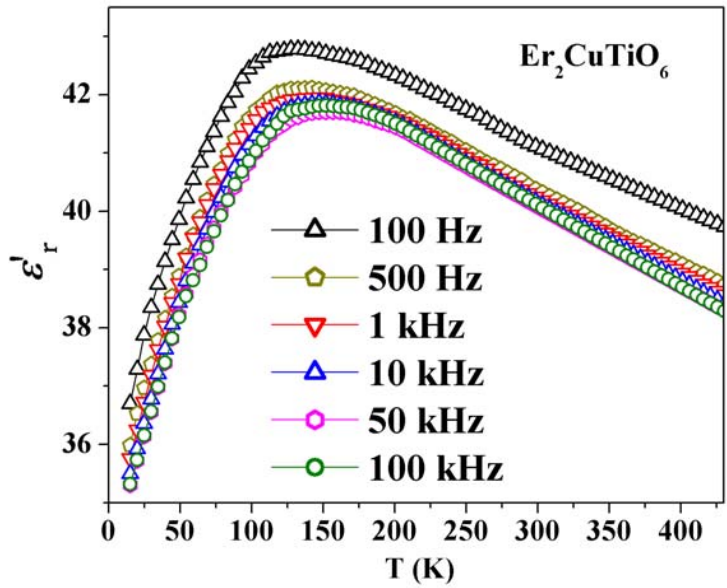

Figure 3: (Color Online) Temperature dependence of $\varepsilon'_r$ of $Er_2CuTiO_6$ at various frequencies.

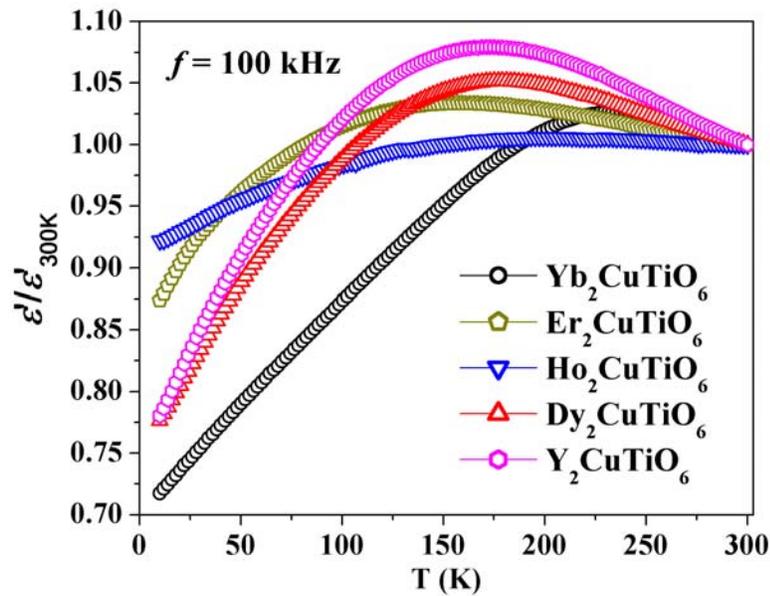

Figure 4: (Color Online) Temperature dependence of the dielectric constant values for various $Ln_2CuTiO_6$ compounds ($Ln$=Y, Ho, Yb, Er, Dy) (normalized with respect to respective dielectric constant values at 300 K) at an applied frequency of 100 kHz.

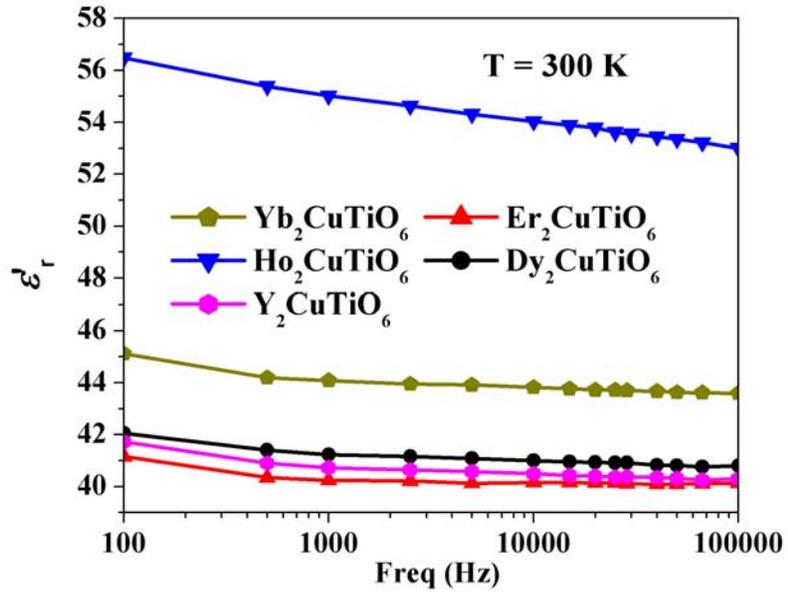

Figure 5: (Color Online) Frequency dependencies of dielectric constant values of various $Ln_2CuTiO_6$ compounds ($Ln$=Y, Ho, Yb, Er, Dy) at 300 K.

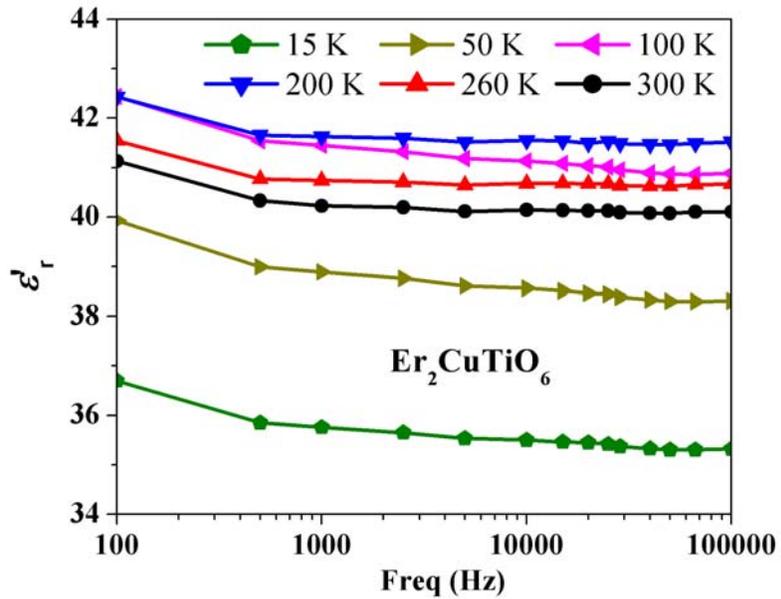

Figure 6: (Color Online) Frequency dependencies of the dielectric constant values for $Er_2CuTiO_6$ for various fixed temperatures.

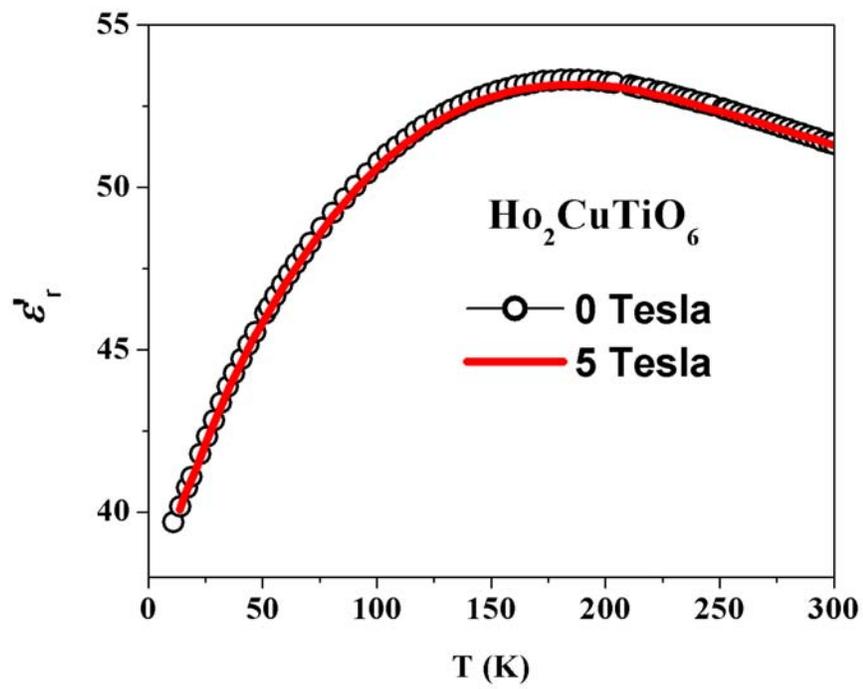

Figure 7: (Color Online) Temperature dependence of dielectric constant of $Ho_2CuTiO_6$ in presence and absence of a magnetic field of 5 Tesla.

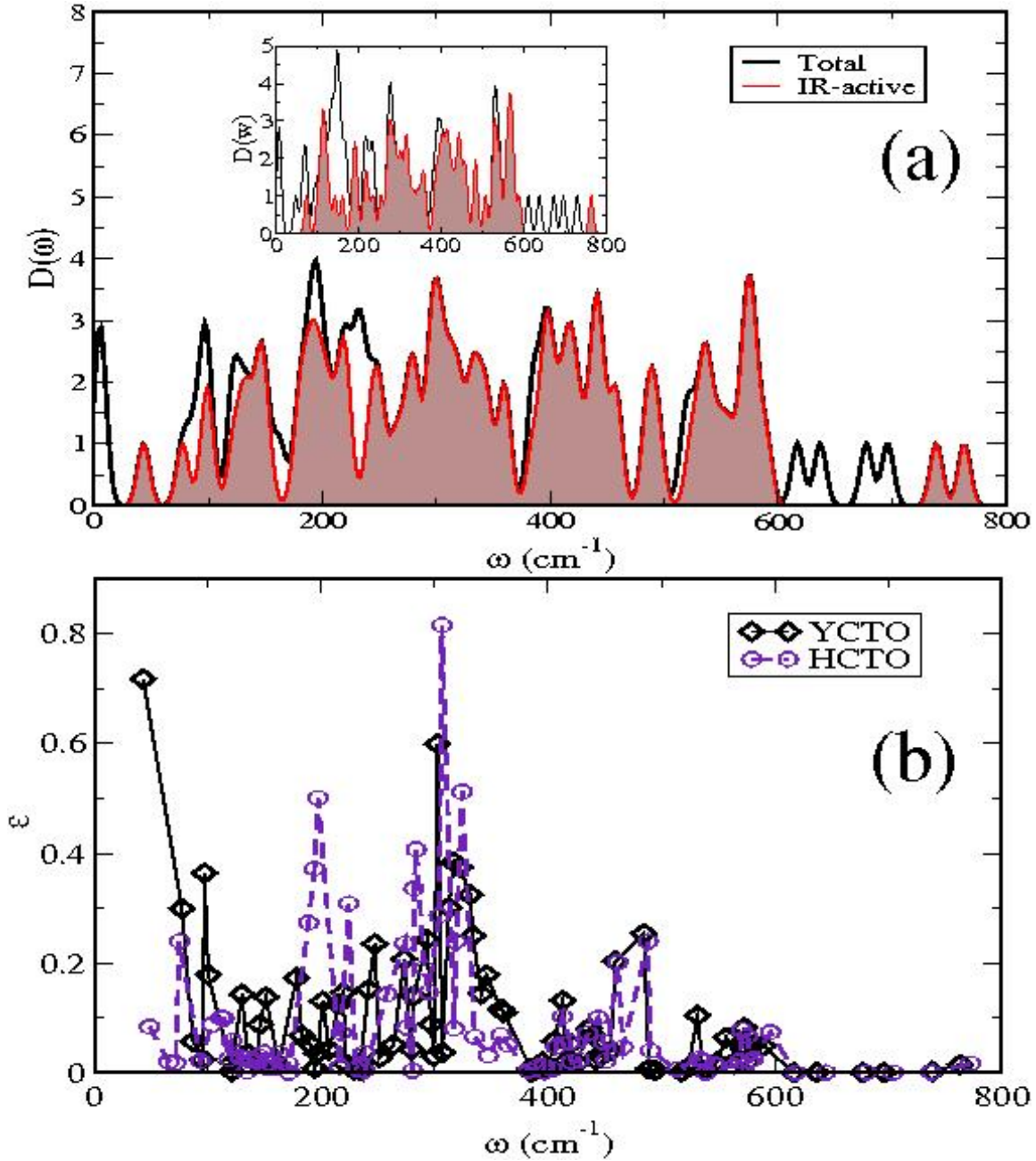

Figure 8: (Color Online) (a) Vibrational density of states of $Y_2CuTiO_6$ at $\Gamma$ point. Inset shows the vibrational density of states of $Ho_2CuTiO_6$ at $\Gamma$ point.
(b) Contribution of phonons to dielectric response of $Y_2CuTiO_6$ (YCTO) and $Ho_2CuTiO_6$ (HCTO), averaged over *x*, *y* and *z* directions. The softest IR-active mode in HCTO contributes a small fraction of the total dielectric response, whereas it is relatively higher in case of YCTO.